\documentclass{PoS}

\usepackage{amssymb}
\usepackage{amsthm}
\usepackage{amsmath}
\usepackage{booktabs}
\usepackage{bm}
\usepackage{bbm}
\DeclareMathOperator{\tr}{tr}
\DeclareMathOperator{\Tr}{Tr}

\newcommand{\Slash}[1]{{\ooalign{\hfil/\hfil\crcr$#1$}}}
\numberwithin{equation}{section}
\usepackage{url}

\title{Energy--momentum tensor on the lattice: recent developments%
}

\ShortTitle{Energy--momentum tensor on the lattice: recent developments}

\author{\speaker{Hiroshi Suzuki}
\\
Department of Physics, Kyushu University, 744 Motooka, Nishi-ku, Fukuoka,
819-0395, Japan\\
        E-mail: \email{hsuzuki@phys.kyushu-u.ac.jp}}

\abstract{It is conceivable that the construction of the energy--momentum
tensor (EMT) in lattice field theory enlarges our ability in lattice field
theory and also deepens our understanding on EMT at the non-pertubative level.
In this talk, I will review recent developments in this enterprise.}

\FullConference{34th annual International Symposium on Lattice Field Theory\\
		24-30 July 2016\\
		University of Southampton, UK}

\begin{document}

\section{Introduction}
\label{sec:1}
The energy--momentum tensor (EMT) is a fundamental observable in quantum field
theory, being the Noether current associated with the translational invariance.
It generates the Poincar\'e transformations and the dilatation through the
Ward--Takahashi (WT) relations. It thus carries information concerning physical
quantities associated with those spacetime symmetries, such as the energy,
momentum, pressure, stress, angular momentum, viscosity, specific heat, and
renormalization group functions. It is also a source of gravity. Thus, we have
enough motivation to compute EMT by employing the lattice regularization, a
most well-established non-perturbative formulation of quantum field theory. It
has been recognized that, however, the construction of EMT on the lattice is
not straightforward because the lattice regularization explicitly breaks the
translational invariance; roughly speaking the situation is similar to the
chiral symmetry on the lattice. In the present talk, I will review recent
developments made on the construction of EMT in lattice (gauge) field theory.

\section{EMT with the dimensional regularization}
\label{sec:2}
First, let us note that the description of EMT in gauge theory is particularly
simple if one adopts the dimensional regularization. This is because this
regularization manifestly preserves the (vectorial) gauge symmetry and the
translational invariance. One can then derive WT relations associated with the
translational invariance straightforwardly. For simplicity, let us take the
pure Yang--Mills theory in a $D$-dimensional Euclidean spacetime,
\begin{equation}
   S=\frac{1}{4g_0^2}\int d^Dx\,F_{\mu\nu}^a(x)F_{\mu\nu}^a(x),
\label{eq:(2.1)}
\end{equation}
where $g_0$ is the bare gauge coupling and
\begin{equation}
   F_{\mu\nu}(x)
   =\partial_\mu A_\nu(x)-\partial_\nu A_\mu(x)+[A_\mu(x),A_\nu(x)]
\label{eq:(2.2)}
\end{equation}
is the field strength. Under the infinitesimal variation,
\begin{equation}
   \delta_\alpha A_\mu(x)=\alpha_\nu(x)F_{\nu\mu}(x),
\label{eq:(2.3)}
\end{equation}
which is a particular combination of the translation with a localized
parameter~$\alpha_\nu(x)$ and a gauge transformation, the action changes as
\begin{equation}
   \delta_\alpha S
   =-\int d^Dx\,\alpha_\nu(x)\partial_\mu T_{\mu\nu}(x),
\label{eq:(2.4)}
\end{equation}
where $T_{\mu\nu}(x)$ is EMT in the pure Yang--Mills theory:
\begin{equation}
   T_{\mu\nu}(x)
   =\frac{1}{g_0^2}\left[
   F_{\mu\rho}^a(x)F_{\nu\rho}^a(x)
   -\frac{1}{4}\delta_{\mu\nu}F_{\rho\sigma}^a(x)F_{\rho\sigma}^a(x)
   \right].
\label{eq:(2.5)}
\end{equation}
Thus, considering the change of variable of the form~\eqref{eq:(2.3)} in the
functional integral containing gauge invariant operators $\mathcal{O}(y)$
and~$\mathcal{O}(z)$, we have\footnote{Here, we implicitly assume the existence
of the gauge-fixing and the Faddeev--Popov ghost terms. The variation of these
terms, however, is BRS exact and can be neglected in correlation functions of
gauge invariant operators.}
\begin{align}
   \left\langle
   \partial_\mu T_{\mu\nu}(x)\mathcal{O}(y)\mathcal{O}(z)
   \right\rangle
   &=-\left\langle\delta(x-y)\partial_\nu\mathcal{O}(y)\mathcal{O}(z)
   +\mathcal{O}(y)\delta(x-z)\partial_\nu\mathcal{O}(z)\right\rangle
\notag\\
   &\qquad{}
   +(\text{terms being vanishing under the integration over~$x$}),
\label{eq:(2.6)}
\end{align}
(the last term can have the structure such as~$\partial_\nu\delta(x-y)$). This
WT relation shows that EMT generates the infinitesimal translation and, at the
same time, the bare quantity $\partial_\mu T_{\mu\nu}(x)$ does not receive any
multiplicative renormalization. In the pure Yang--Mills theory, there is no
gauge invariant second-rank symmetric tensor of the mass dimension smaller than
or equal to~$4$ whose total divergence identically vanishes; the only exception
is a constant multiple of~$\delta_{\mu\nu}$. Thus the absolute normalization of
EMT is fixed by the physical requirement~\eqref{eq:(2.6)}. The renormalized
finite EMT is then obtained by subtracting a possibly divergent vacuum
expectation value (VEV),\footnote{With the dimensional regularization,
$T_{\mu\nu}$ will be finite even without such a subtraction; here we define the
renormalized EMT by subtracting VEV to make expressions be consistent also with
the lattice regularization.}
\begin{equation}
   T_{\mu\nu}(x)\to T_{\mu\nu}(x)-\left\langle T_{\mu\nu}(x)\right\rangle.
\label{eq:(2.7)}
\end{equation}
Thus the construction of EMT is particularly simple with the dimensional
regularization. This is the reason why the dimensional regularization is
extensively employed to study the issue of the trace or conformal anomaly in
gauge theory~\cite{Collins:1976yq}, for example. The drawback of the
dimensional regularization is of course that it is defined only within
perturbation theory.

\section{EMT with the lattice regularization}
\label{sec:3}
A general strategy for the construction of EMT in lattice field theory was
developed in detail in~Refs.~\cite{Caracciolo:1989pt,Caracciolo:1991cp}. The
strategy is quite analogous to the one for the chiral symmetry on the
lattice studied in~Ref.~\cite{Bochicchio:1985xa}.\footnote{It is interesting
to note that a computation of the trace anomaly of the Wilson fermion in a
pioneering work~\cite{Fujikawa:1983az} yielded $-15$ times the conventional
trace anomaly. Later, it was realized in~Ref.~\cite{Caracciolo:1989pt} that
the reason for this discrepancy is that the translation WT relation is not
imposed in~Ref.~\cite{Fujikawa:1983az}.}

We start with a possible definition of a ``would-be infinitesimal translation''
on the link variables~$U(x,\mu)$. Since the infinitesimal translation is in any
case broken by the lattice structure, the definition is to a large extent
arbitrary. Here, corresponding to~Eq.~\eqref{eq:(2.3)}, we set
\begin{equation}
   \Hat{\delta}_\alpha U(x,\mu)=\alpha_\nu(x)\Hat{F}_{\nu\mu}(x)U(x,\mu).
\label{eq:(3.1)}
\end{equation}
In principle, we can use any lattice transcription for the field strength,
$\Hat{F}_{\nu\mu}(x)$. To make the transformation under the hypercubic
transformation simple, however, it is convenient to adopt the clover definition
for~$\Hat{F}_{\nu\mu}(x)$. We then consider the change of variable of the
form~\eqref{eq:(3.1)} in the functional integral containing gauge invariant
lattice operators~$\Hat{\mathcal{O}}(y)$ and~$\Hat{\mathcal{O}}(z)$. This time,
we have ($\Hat{\partial}_\mu$ being a lattice derivative)
\begin{equation}
   \left\langle
   \left[\Hat{\partial}_\mu\Hat{T}_{\mu\nu}^{\text{naive}}(x)
   +\Hat{X}_\nu(x)\right]\Hat{\mathcal{O}}(y)\Hat{\mathcal{O}}(z)\right\rangle
   =-\left\langle\frac{\delta}{\delta\alpha_\nu(x)}
   \Hat{\delta}_\alpha\Hat{\mathcal{O}}(y)\Hat{\mathcal{O}}(z)
   +\Hat{\mathcal{O}}(y)\frac{\delta}{\delta\alpha_\nu(x)}
   \Hat{\delta}_\alpha\Hat{\mathcal{O}}(z)\right\rangle.
\label{eq:(3.2)}
\end{equation}
In this expression, the naive lattice transcription of the classical EMT
$\Hat{T}_{\mu\nu}^{\text{naive}}(x)$ is defined by
\begin{equation}
   \Hat{T}_{\mu\nu}^{\text{naive}}(x)
   \equiv\Hat{T}_{\mu\nu}^{[6]}(x)+\Hat{T}_{\mu\nu}^{[3]}(x)
\label{eq:(3.3)}
\end{equation}
from
\begin{align}
   \Hat{T}_{\mu\nu}^{[6]}(x)&\equiv\frac{1}{g_0^2}(1-\delta_{\mu\nu})
   \sum_\rho\Hat{F}_{\mu\rho}^a(x)\Hat{F}_{\nu\rho}^a(x),
\label{eq:(3.4)}
\\
   \Hat{T}_{\mu\nu}^{[3]}(x)&\equiv\frac{1}{g_0^2}
   \delta_{\mu\nu}\left[
   \sum_\rho\Hat{F}_{\mu\rho}^a(x)\Hat{F}_{\nu\rho}^a(x)
   -\frac{1}{4}\sum_{\rho\sigma}\Hat{F}_{\rho\sigma}^a(x)\Hat{F}_{\rho\sigma}^a(x)
   \right].
\label{eq:(3.5)}
\end{align}
These $\Hat{T}_{\mu\nu}^{[6]}(x)$ and~$\Hat{T}_{\mu\nu}^{[3]}(x)$ transform as the
sextet and the triplet representations under the hypercubic group,
respectively; $\Hat{T}_{\mu\nu}^{[6]}(x)$ possesses \emph{only off-diagonal
components} in~$\mu$ and~$\nu$, while $\Hat{T}_{\mu\nu}^{[3]}(x)$ possesses
\emph{only diagonal components and is traceless}. Since
$\Hat{T}_{\mu\nu}^{\text{naive}}(x)$ coincides with EMT in the classical theory in
the tree-level continuum limit, the breaking term~$\Hat{X}_\nu(x)$
in~Eq.~\eqref{eq:(3.2)}, which arises from the non-invariance of the lattice
action under the would-be translation~\eqref{eq:(3.1)}
with~$\alpha_\nu=\text{const.}$, is $O(a)$ in the tree-level continuum limit.
However, since $\Hat{T}_{\mu\nu}^{\text{naive}}(x)$ is traceless
$\sum_\mu\Hat{T}_{\mu\mu}^{\text{naive}}(x)=0$ and does not reproduce the trace
anomaly~\cite{Collins:1976yq}, $\Hat{T}_{\mu\nu}^{\text{naive}}(x)$ cannot be the
correct EMT in quantum theory.

In fact, though the radiative corrections, the breaking term~$\Hat{X}_\nu(x)$
can be $O(a^0)$ in correlation functions. Assuming lattice symmetries such as
the hypercubic symmetry, one infers the expansion,
\begin{equation}
   \Hat{X}_\nu(x)
   =\left(\frac{Z_6}{Z_\delta}-1\right)\Hat{\partial}_\mu\Hat{T}_{\mu\nu}^{[6]}(x)
   +\left(\frac{Z_3}{Z_\delta}-1\right)\Hat{\partial}_\mu
   \Hat{T}_{\mu\nu}^{[3]}(x)
   +\frac{Z_1}{Z_\delta}\Hat{\partial}_\mu\Hat{T}_{\mu\nu}^{[1]}(x)
   +\frac{1}{Z_\delta}\Hat{R}_\nu(x),
\label{eq:(3.6)}
\end{equation}
where $Z_\alpha$ ($\alpha=6$, $3$, and~$1$) and~$Z_\delta$ are constants and the
last term~$\Hat{R}_\nu(x)$ is $O(a)$ in correlation functions if it is
separated from other operators in position space. In~Eq.~\eqref{eq:(3.6)}, the
new combination
\begin{equation}
   \Hat{T}_{\mu\nu}^{[1]}(x)\equiv\frac{1}{g_0^2}\delta_{\mu\nu}\sum_{\rho\sigma}
   \Hat{F}_{\rho\sigma}^a(x)\Hat{F}_{\rho\sigma}^a(x)
\label{eq:(3.7)}
\end{equation}
transforms as the singlet representation under the hypercubic group;
$\Hat{T}_{\mu\nu}^{[1]}(x)$ possesses \emph{only diagonal components and is
trace-ful}.

Substituting the expansion~\eqref{eq:(3.6)} into the identity~\eqref{eq:(3.2)},
we have
\begin{align}
   &\left\langle
   \Hat{\partial}_\mu\left[
   Z_6\Hat{T}_{\mu\nu}^{[6]}(x)
   +Z_3\Hat{T}_{\mu\nu}^{[3]}(x)+Z_1\Hat{T}_{\mu\nu}^{[1]}(x)
   \right]
   \Hat{\mathcal{O}}(y)\Hat{\mathcal{O}}(z)\right\rangle
\notag\\
   &=-\left\langle Z_\delta\frac{\delta}{\delta\alpha_\nu(x)}
   \Hat{\delta}_\alpha\Hat{\mathcal{O}}(y)\Hat{\mathcal{O}}(z)
   +\Hat{\mathcal{O}}(y)Z_\delta\frac{\delta}{\delta\alpha_\nu(x)}
   \Hat{\delta}_\alpha\Hat{\mathcal{O}}(z)
   +\Hat{R}_\nu(x)\Hat{\mathcal{\mathcal{O}}}(y)\Hat{\mathcal{\mathcal{O}}}(z)
   \right\rangle.
\label{eq:(3.8)}
\end{align}
The statement of the restoration of the translational invariance in the
continuum limit is that one can choose the constant~$Z_\delta$ as~$a\to0$ so
that the right-hand side of the above expression becomes (recall
Eq.~\eqref{eq:(2.6)})
\begin{align}
   &\stackrel{a\to0}{\to}
   -\left\langle\delta(x-y)\partial_\nu\mathcal{O}(y)\mathcal{O}(z)
   +\mathcal{O}(y)\delta(x-z)\partial_\nu\mathcal{O}(z)\right\rangle
\notag\\
   &\qquad\qquad{}
   +(\text{terms being vanishing under the integration over~$x$}),
\label{eq:(3.9)}
\end{align}
where $\mathcal{O}(y)$ and $\mathcal{O}(z)$ are operators in the continuum
theory corresponding to $\Hat{\mathcal{O}}(y)$ and $\Hat{\mathcal{O}}(z)$,
respectively. Assuming this is possible, the properly-normalized EMT which
reproduces the correct WT relation in the continuum limit is given by
\begin{equation}
   \Hat{T}_{\mu\nu}(x)
   =Z_6\Hat{T}_{\mu\nu}^{[6]}(x)+Z_3\Hat{T}_{\mu\nu}^{[3]}(x)
   +Z_1\left[\Hat{T}_{\mu\nu}^{[1]}(x)
   -\left\langle\Hat{T}_{\mu\nu}^{[1]}(x)\right\rangle\right],
\label{eq:(3.10)}
\end{equation}
where we have subtracted the vacuum expectation value (only the trace-ful
diagonal~$\Hat{T}_{\mu\nu}^{[1]}(x)$ can possess the vacuum expectation value).
Thus, our mission is to determine the renormalization constants, $Z_6$, $Z_3$,
and~$Z_1$.

There are various strategies to determine the renormalization constants:
\begin{itemize}
\item Eqs.~\eqref{eq:(3.8)} and~\eqref{eq:(3.9)} reduce to the conservation law
of EMT if we set $x\neq y$ and~$x\neq z$. This determines the ratios $Z_3/Z_6$
and~$Z_1/Z_6$. The overall normalization~$Z_6$ in~Eq.~\eqref{eq:(3.10)} may be
fixed by the rest energy $-\Hat{T}_{00}$ of a hadronic
state~\cite{Caracciolo:1989pt,Caracciolo:1991cp}.
\item Use a matching of the thermal expectation value of~Eq.~\eqref{eq:(3.10)}
with bulk thermodynamic quantities (the energy-density and the pressure)
obtained by one of standard methods~\cite{Meyer:2007ic,Huebner:2008as}. This
determines $Z_3$ and~$Z_1$ but tells nothing about~$Z_6$, the coefficient of
off-diagonal elements of EMT.
\item Use physical normalization conditions implied by shifted boundary
conditions; $Z_6$ can also be obtained~\cite{Giusti:2010bb,Giusti:2011kt,%
Giusti:2012yj,Robaina:2013zmb,Giusti:2014ila,Giusti:2014tfa,Giusti:2015daa,%
Giusti:2015got}.
\item Use the gradient flow~\cite{Narayanan:2006rf,Luscher:2009eq,%
Luscher:2010iy,Luscher:2011bx} for the probe
operators~$\Hat{\mathcal{O}}(y)\Hat{\mathcal{O}}(z)$
in~Eq.~\eqref{eq:(3.8)}~\cite{DelDebbio:2013zaa,Patella:2014dsa,%
Capponi:2015ucc,Capponi:2015ahp}.
\end{itemize}
In the present talk, I will review the last two recent ideas.
In~Sect.~\ref{sec:6}, I will explain an alternative approach which does not
rely on the representation~\eqref{eq:(3.10)}; it employs the gradient flow to
construct EMT itself~\cite{Suzuki:2013gza,Asakawa:2013laa,Makino:2014taa,%
Makino:2014sta,Kitazawa:2014uxa,Makino:2014cxa,Suzuki:2015fka,Itou:2015gxx,%
Kitazawa:2015ewp,Taniguchi:2016ofw,Kitazawa:2016dsl,Kanaya:2016rkt}.

\section{Shifted boundary conditions~\cite{Giusti:2010bb,%
Giusti:2011kt,Giusti:2012yj,Robaina:2013zmb,Giusti:2014ila,%
Giusti:2014tfa,Giusti:2015daa,Giusti:2015got}}

By employing boundary conditions with a spatial shift, one can derive physical
normalization conditions which can be used to determine renormalization
constants for EMT in~Eq.~\eqref{eq:(3.10)}, especially $Z_6$ for the
off-diagonal elements. For this, we consider the free energy density with a
spatial shift vector~$\bm{\xi}$:
\begin{equation}
   f(L_0,\bm{\xi})
   =-\frac{1}{V}\ln\Tr\left[e^{-L_0(H-i\bm{\xi}\cdot\bm{P})}\right],
\label{eq:(4.1)}
\end{equation}
where $L_0$ denotes the temporal size of the system and $V$ is the
spatial volume; $H$ and~$\bm{P}$ are the Hamiltonian and the momentum
operators, respectively. This shifted partition function can be expressed by the
functional integral over a field~$\phi(x_0,\bm{x})$ with the shifted boundary
condition, $\phi(L_0,\bm{x})=\phi(0,\bm{x}-L_0\bm{\xi})$. Then the derivative
with respect to the $i$-th component of the shift vector, $\xi_i$, gives rise
to the $i$-th momentum,
\begin{equation}
   \frac{\partial}{\partial\xi_i}f(L_0,\bm{\xi})
   =-\frac{1}{V}iL_0\left\langle P_i\right\rangle_{\bm{\xi}}
   =-L_0\left\langle T_{0i}\right\rangle_{\bm{\xi}}.
\label{eq:(4.2)}
\end{equation}
This condition can be used to determine $Z_6$ in~Eq.~\eqref{eq:(3.10)} if the
left-hand side of this relation can be computed; for an efficient method for
this computation, see~Ref.~\cite{Giusti:2015daa}.

We can also derive interesting relations by relating systems with shifted and
un-shifted boundary conditions respectively by noting the underlying $SO(4)$
symmetry (in the continuum theory). For a shift vector of the
form~$\bm{\xi}=(\xi_1,0,0)$, for example, the $SO(4)$ rotation
\begin{equation}
   R=\begin{pmatrix}
   \gamma_1&\gamma_1\xi_1&0&0\\
   -\gamma_1\xi_1&\gamma_1&0&0\\
   0&0&1&0\\
   0&0&0&1\\
   \end{pmatrix},\qquad
   \gamma_1=\frac{1}{\sqrt{1+\xi_1^2}},
\label{eq:(4.3)}
\end{equation}
make the boundary condition un-shifted. Since EMT transforms as a tensor under
the rotation, we have
\begin{equation}
   \left\langle T_{\mu\nu}\right\rangle_{\bm{\xi}}
   =\sum_{\rho,\sigma}R^T_{\mu\rho}R^T_{\nu\sigma}
   \left\langle T_{\rho\sigma}\right\rangle,
\label{eq:(4.4)}
\end{equation}
where the expectation value in the right-hand side is defined with respect to
the boundary condition without the shift. By considering this relation for
$(\mu,\nu)=(0,0)$, $(0,1)$, and~$(1,1)$ and noting
$\langle T_{01}\rangle=0$ in the thermodynamic limit (which is assumed here),
we have
\begin{equation}
   \left\langle T_{0i}\right\rangle_{\bm{\xi}}
   =\frac{\xi_i}{1-\xi_i^2}
   \left\langle T_{00}-T_{ii}\right\rangle_{\bm{\xi}},
\label{eq:(4.5)}
\end{equation}
by eliminating the expectation values with the un-shifted boundary condition
from the relation~\eqref{eq:(4.4)} in favor of the expectation values with the
shifted one. This relation can be used to determine the ratio $Z_3/Z_6$.

One can also derive
\begin{equation}
   \frac{\partial}{\partial\xi_i}
   \left\langle\sum_\mu T_{\mu\mu}\right\rangle_{\bm{\xi}}
   =\frac{1}{(1+\bm{\xi}^2)^2}\frac{\partial}{\partial\xi_i}
   \left[
   \frac{(1+\bm{\xi}^2)^3}{\xi_i}
   \left\langle T_{0i}\right\rangle_{\bm{\xi}}\right],
\label{eq:(4.6)}
\end{equation}
which may be used to determine the ratio~$Z_1/Z_6$. 


On the basis of the renormalization constants $Z_6$ and~$Z_3/Z_6$ obtained from
the above relations, a very accurate entropy density of the $SU(3)$ pure
Yang--Mills theory at finite temperature is calculated~\cite{Giusti:2015got}.
See also Ref.~\cite{Giusti:2016wsf} for updated results.

\section{The gradient flow is used for the probe
operators~\cite{DelDebbio:2013zaa,Patella:2014dsa,Capponi:2015ucc,%
Capponi:2015ahp}}

Let us recall the would-be translation WT relation, Eq.~\eqref{eq:(3.8)}. In
the right-hand side of this equation, the operator~$\Hat{R}_\nu(x)$ can provide
$O(a^0)$ contributions when it collides with other operators (the so-called
contact term) although $\Hat{R}_\nu(x)$ itself is $O(a)$ when it is separated
from other operators in position space. This can happen, in particular, when we
integrate the WT relation over the coordinate~$x$; the point~$x$ then collides
with the points $y$ and~$z$. Thus, it is not straightforward to isolate the
terms being proportional to~$Z_\delta$ in~Eq.~\eqref{eq:(3.8)}. As we will see
below, however, the gradient flow~\cite{Narayanan:2006rf,Luscher:2009eq,%
Luscher:2010iy,Luscher:2011bx} provides an interesting method to obtain
$Z_\delta$ from the requirement~\eqref{eq:(3.8)} and at the same time $Z_6$,
$Z_3$, and~$Z_1$ from the WT relation.

The gradient flow is a one-parameter evolution of the gauge field. For the
gauge potential in the continuum theory~$A_\mu(x)$, the evolution along the
flow time~$t$ is defined by
\begin{equation}
   \partial_tB_\mu(t,x)=D_\nu G_{\nu\mu}(t,x),\qquad B_\mu(t=0,x)=A_\mu(x),
\label{eq:(5.1)}
\end{equation}
where
\begin{align}
   G_{\mu\nu}(t,x)
   &=\partial_\mu B_\nu(t,x)-\partial_\nu B_\mu(t,x)
   +[B_\mu(t,x),B_\nu(t,x)],
\label{eq:(5.2)}
\\
   D_\mu&=\partial_\mu+[B_\mu,\cdot],
\label{eq:(5.3)}
\end{align}
are the field strength of the flowed gauge field and the covariant derivative
with respect to~$B_\mu(t,x)$, respectively. I emphasize that the initial
condition of the flow equation~\eqref{eq:(5.1)} is the bare gauge field. A
surprising feature of the gradient flow, which was perturbatively proven
in~Ref.~\cite{Luscher:2011bx} (see also Ref.~\cite{Hieda:2016xpq}), is that any
composite operator of the flowed gauge field for a positive flow time~$t>0$
automatically becomes a renormalized finite operator; moreover it does not
produce any new UV divergences even if other composite operators collide with
it.

Thus let us suppose that we take certain composite operators composed from the
flowed gauge field as the probe
operators~$\Hat{\mathcal{O}}(y)\Hat{\mathcal{O}}(z)$ in~Eq.~\eqref{eq:(3.8)}.
For example, we may take
\begin{equation}
   \Hat{\mathcal{O}}(y)
   \equiv\left.\Hat{\partial}_\rho
   \Hat{T}_{\rho\nu}^{[\alpha]}(y)
   \right|_{\text{flowed lattice gauge field at the flow time~$t>0$}},\qquad
   \alpha=6,3,1,\qquad
   \Hat{\mathcal{O}}(z)\equiv1.
\label{eq:(5.4)}
\end{equation}
Then because of the UV finiteness of the gradient flow, as~$a\to0$ the
contribution of~$\Hat{R}_\nu(x)$ in~Eq.~\eqref{eq:(3.8)} can be neglected
even for~$x=y$ and $x=z$ and we have\footnote{Here, we adopt a boundary
condition which breaks the translational invariance; then generally the
right-hand side does not vanish.}
\begin{align}
   &\left\langle
   \Hat{\partial}_\mu\left[
   Z_6\Hat{T}_{\mu\nu}^{[6]}(x)
   +Z_3\Hat{T}_{\mu\nu}^{[3]}(x)+Z_1\Hat{T}_{\mu\nu}^{[1]}(x)
   \right]
   \Hat{\mathcal{O}}(y)\right\rangle
   =-\left\langle Z_\delta\frac{\delta}{\delta\alpha_\nu(x)}
   \Hat{\delta}_\alpha\Hat{\mathcal{O}}(y)
   \right\rangle+O(a)
\notag\\
   &\stackrel{a\to0}{\to}
   -\left\langle\delta(x-y)\partial_\nu\mathcal{O}(y)
   \right\rangle
   +(\text{terms being vanishing under the integration over~$x$}).
\label{eq:(5.5)}
\end{align}
Since we know the explicit form of~$\frac{\delta}{\delta\alpha_\nu(x)}
\Hat{\delta}_\alpha\Hat{\mathcal{O}}(y)$, the first equality for the prove
operator~\eqref{eq:(5.4)} with~$\alpha=6$, $3$, and~$1$, provides a possible
method to determine the ratios $Z_6/Z_\delta$, $Z_3/Z_\delta$, and $Z_1/Z_\delta$
for~$a\to0$. Also from the last requirement in the continuum limit, one may
extract the renormalization constant~$Z_\delta$ itself. For explicit results
obtained by the above strategy, see Refs.~\cite{Capponi:2015ahp}
and~\cite{Capponi}. For an application to the 3D $\lambda\phi^4$~theory,
see~Ref.~\cite{Capponi:2016yjz}.

\section{Universal formula for EMT~\cite{Suzuki:2013gza,Asakawa:2013laa,%
Makino:2014taa,Makino:2014sta,Kitazawa:2014uxa,Makino:2014cxa,Suzuki:2015fka,%
Itou:2015gxx,Kitazawa:2015ewp,Taniguchi:2016ofw,Kitazawa:2016dsl,%
Kanaya:2016rkt}}
\label{sec:6}
\subsection{Small flow time representation of EMT}
In this section, I explain an alternative approach~\cite{Suzuki:2013gza,%
Makino:2014taa} to EMT on the lattice, which is based on the UV finiteness of
composite operators constructed by the gradient flow~\eqref{eq:(5.1)} and
the flow for the fermion field~\cite{Luscher:2013cpa}:
\begin{align}
   \partial_t\chi(t,x)&=\Delta\chi(t,x),&
   \chi(t=0,x)&=\psi(x),
\label{eq:(6.1)}\\
   \partial_t\Bar{\chi}(t,x)
   &=\Bar{\chi}(t,x)
   \overleftarrow{\Delta},
   &\Bar{\chi}(t=0,x)&=\Bar{\psi}(x),
\label{eq:(6.2)}
\end{align}
where, on fermion fields,
\begin{align}
   \Delta&=D_\mu D_\mu,&D_\mu&=\partial_\mu+B_\mu,
\label{eq:(6.3)}\\
   \overleftarrow{\Delta}&\equiv\overleftarrow{D}_\mu\overleftarrow{D}_\mu,
   &\overleftarrow{D}_\mu&\equiv\overleftarrow{\partial}_\mu-B_\mu.
\label{eq:(6.4)}
\end{align}

This approach gives rise to a ``universal formula'' for EMT, which is rather
different from the representation~\eqref{eq:(3.10)}.

The approach can be developed for general vector-like gauge theories
($D_\mu\equiv\partial_\mu+A_\mu$ for the fermion):
\begin{equation}
   S=\frac{1}{4g_0^2}\int d^Dx\,F_{\mu\nu}^a(x)F_{\mu\nu}^a(x)
   +\int d^Dx\,\Bar{\psi}(x)(\Slash{D}+m_0)\psi(x),
\label{eq:(6.5)}
\end{equation}
where $m_0$ is the bare fermion mass parameter. As in Sect.~\ref{sec:2},
assuming the dimensional regularization, by considering an infinitesimal
variation of integration variables in the functional integral of the form,
\begin{equation}
   \delta_\alpha A_\mu(x)=\alpha_\nu(x)F_{\nu\mu}(x),\qquad
   \delta_\alpha\psi(x)=\alpha_\mu(x)D_\mu\psi(x),
\label{eq:(6.6)}
\end{equation}
we have EMT which satisfies the translation WT identity. The explicit form is
given by~\cite{Collins:1976yq}
\begin{align}
   T_{\mu\nu}(x)
   &=\frac{1}{g_0^2}\left[
   F_{\mu\rho}^a(x)F_{\nu\rho}^a(x)
   -\frac{1}{4}\delta_{\mu\nu}F_{\rho\sigma}^a(x)F_{\rho\sigma}^a(x)
   \right]
\notag\\
   &\qquad{}
   +\frac{1}{4}
   \Bar{\psi}(x)\left(\gamma_\mu\overleftrightarrow{D}_\nu
   +\gamma_\nu\overleftrightarrow{D}_\mu\right)\psi(x)
   -\delta_{\mu\nu}\Bar{\psi}(x)
   \left(\frac{1}{2}\overleftrightarrow{\Slash{D}}
   +m_0\right)\psi(x),
\label{eq:(6.7)}
\end{align}
where
\begin{equation}
   \overleftrightarrow{D}_\mu\equiv D_\mu-\overleftarrow{D}_\mu,\qquad
   \overleftarrow{D}_\mu\equiv\overleftarrow{\partial}_\mu-A_\mu.
\label{eq:(6.8)}
\end{equation}
Note however that the expression~\eqref{eq:(6.7)} is meaningful only under the
perturbative dimensional regularization; it cannot be employed with the lattice
regularization as it stands.

At this point we recall that any local composite operator composed from the
flowed gauge field $B_\mu(t,x)$ in~Eq.~\eqref{eq:(5.1)} and the flowed fermion
fields $\chi(t,x)$ in~Eq.~\eqref{eq:(6.1)} and $\Bar{\chi}(t,x)$
in~Eq.~\eqref{eq:(6.2)} for~$t>0$ automatically becomes a renormalized finite
operator, if the flowed fermion fields are multiplicatively
renormalized~\cite{Luscher:2011bx,Luscher:2013cpa} (see
also~Ref.~\cite{Hieda:2016xpq}):
\begin{equation}
   \chi_R(t,x)=Z_\chi^{1/2}\chi(t,x),\qquad
   \Bar{\chi}_R(t,x)=Z_\chi^{1/2}\Bar{\chi}(t,x).
\end{equation}
These wave function renormalizations may be avoided by using the following
``ringed variables''~\cite{Makino:2014taa}:\footnote{In what follows, we
assume that all fermion masses are degenerate; $N_f$ denotes the number of
degenerated flavors.}
\begin{align}
   \mathring{\chi}(t,x)
   &\equiv\sqrt{\frac{-2\dim(R)N_f}
   {(4\pi)^2t^2
   \left\langle\Bar{\chi}(t,x)\overleftrightarrow{\Slash{D}}\chi(t,x)
   \right\rangle}}
   \,\chi(t,x),
\label{eq:(6.9)}\\
   \mathring{\Bar{\chi}}(t,x)
   &\equiv\sqrt{\frac{-2\dim(R)N_f}
   {(4\pi)^2t^2
   \left\langle\Bar{\chi}(t,x)\overleftrightarrow{\Slash{D}}\chi(t,x)
   \right\rangle}}
   \,\Bar{\chi}(t,x),
\label{eq:(6.10)}
\end{align}
where $\dim(R)$ denotes the dimension of the gauge representation~$R$ to which
the fermion is belonging, because the wave function renormalization
constant~$Z_\chi$ is canceled out in~$\mathring{\chi}(t,x)$ and
in~$\mathring{\Bar{\chi}}(t,x)$. Thus any composite of $B_\mu(t,x)$,
$\mathring{\chi}(t,x)$, and~$\mathring{\Bar{\chi}}(t,x)$ for~$t>0$ is a
renormalized finite operator. Such a renormalized finite operator should
possess a meaning being independent of the regularization, if the
renormalization conditions are taken equally. In what follows, we construct a
certain local combination of $B_\mu(t,x)$, $\mathring{\chi}(t,x)$,
and~$\mathring{\Bar{\chi}}(t,x)$ which coincides with~Eq.~\eqref{eq:(6.7)} if
one uses the dimensional regularization. This composite operator of flowed
fields thus becomes EMT under the dimensional regularization and on the other
hand it possesses a meaning being independent of the regularization. Thus,
assuming the existence of EMT in the non-perturbative level, \emph{this
combination must be EMT}. This is our strategy. From this reasoning it is clear
that our expression for EMT is universal in the sense that it should hold with
any regularization, not only with the lattice regularization.

The problem is however how to obtain such a combination of flowed fields which
reduces to~Eq.~\eqref{eq:(6.7)} under the dimensional regularization. The
relationship between the original fields at the vanishing flow time and the
flowed fields at~$t>0$ is quite non-trivial because the flow equations are
non-linear in field variables.

The relationship however can be tractable at least in one situation, the small
flow time limit~$t\to0$. This is the so-called small flow time
expansion~\cite{Luscher:2011bx} which infers that for~$t\to0$ a local composite
operator of flowed fields at~$t>0$ can be represented by an asymptotic series
of~\emph{local\/} composite operators of un-flowed fields with increasing mass
dimensions. For example, in the pure Yang--Mills theory, we have (here,
$\mathbbm{1}$ denotes the identity operator)
\begin{align}
   G_{\mu\rho}^a(t,x)G_{\nu\rho}^a(t,x)
   &\stackrel{t\to0}{\sim}
   \left\langle G_{\mu\rho}^a(t,x)G_{\nu\rho}^a(t,x)\right\rangle
   \mathbbm{1}
\notag\\
   &\qquad{}
   +\zeta_{11}(t)
   \left[F_{\mu\rho}^a(x)F_{\nu\rho}^a(x)
   -\left\langle F_{\mu\rho}^a(x)F_{\nu\rho}^a(x)
   \right\rangle\right]
\notag\\
   &\qquad\qquad{}
   +\zeta_{12}(t)
   \left[\delta_{\mu\nu}F_{\rho\sigma}^a(x)F_{\rho\sigma}^a(x)
   -\left\langle\delta_{\mu\nu}F_{\rho\sigma}^a(x)F_{\rho\sigma}^a(x)
   \right\rangle\right]
   +O(t).
\label{eq:(6.12)}
\end{align}
Since the flow time possesses the mass dimension~$-2$, the $O(t)$ term in the
above expansion starts from local operators of the mass dimension~$6$. The
expansion coefficients such as $\zeta_{11}(t)$ and~$\zeta_{12}(t)$ in the
dimensional regularization for~$t\to0$, moreover, can be worked out by
perturbation theory if the theory is asymptotically free~\cite{Luscher:2011bx}.
See also Ref.~\cite{Suzuki:2015bqa} for an efficient method for the
computation. Thus, considering the $t\to0$ limit to neglect the $O(t)$ term
in~Eq.~\eqref{eq:(6.12)}, one can express the first line of~Eq.~\eqref{eq:(6.7)}
(in the pure Yang--Mills theory) in terms of the flowed gauge field by
inverting the above relation with respect
to~$F_{\mu\rho}^a(x)F_{\nu\rho}^a(x)-\text{VEV}$ (this is always possible within
perturbation theory).

Carrying out the above procedure,\footnote{This idea has been examined
analytically by using solvable models~\cite{Makino:2014sta,Makino:2014cxa,%
Suzuki:2015fka}.} we have~\cite{Suzuki:2013gza,Makino:2014taa}
\begin{align}
   T_{\mu\nu}(x)
   &=\lim_{t\to0}\biggl\{c_1(t)\left[
   \Tilde{\mathcal{O}}_{1\mu\nu}(t,x)
   -\frac{1}{4}\Tilde{\mathcal{O}}_{2\mu\nu}(t,x)
   \right]
\notag\\
   &\qquad\qquad{}
   +c_2(t)\left[
   \Tilde{\mathcal{O}}_{2\mu\nu}(t,x)
   -\left\langle\Tilde{\mathcal{O}}_{2\mu\nu}(t,x)\right\rangle
   \right]
\notag\\
   &\qquad\qquad\qquad{}
   +c_3(t)\left[
   \Tilde{\mathcal{O}}_{3\mu\nu}(t,x)
   -2\Tilde{\mathcal{O}}_{4\mu\nu}(t,x)
   -\left\langle
   \Tilde{\mathcal{O}}_{3\mu\nu}(t,x)
   -2\Tilde{\mathcal{O}}_{4\mu\nu}(t,x)
   \right\rangle
   \right]
\notag\\
   &\qquad\qquad\qquad\qquad{}
   +c_4(t)\left[
   \Tilde{\mathcal{O}}_{4\mu\nu}(t,x)
   -\left\langle\Tilde{\mathcal{O}}_{4\mu\nu}(t,x)\right\rangle
   \right]
\notag\\
   &\qquad\qquad\qquad\qquad\qquad{}
   +c_5(t)\left[
   \Tilde{\mathcal{O}}_{5\mu\nu}(t,x)
   -\left\langle\Tilde{\mathcal{O}}_{5\mu\nu}(t,x)\right\rangle
   \right]\biggr\},
\label{eq:(6.13)}
\end{align}
where
\begin{align}
   \Tilde{\mathcal{O}}_{1\mu\nu}(t,x)&\equiv
   G_{\mu\rho}^a(t,x)G_{\nu\rho}^a(t,x),
\label{eq:(6.14)}\\
   \Tilde{\mathcal{O}}_{2\mu\nu}(t,x)&\equiv
   \delta_{\mu\nu}G_{\rho\sigma}^a(t,x)G_{\rho\sigma}^a(t,x),
\label{eq:(6.15)}\\
   \Tilde{\mathcal{O}}_{3\mu\nu}(t,x)&\equiv
   \mathring{\Bar{\chi}}(t,x)
   \left(\gamma_\mu\overleftrightarrow{D}_\nu
   +\gamma_\nu\overleftrightarrow{D}_\mu\right)
   \mathring{\chi}(t,x),
\label{eq:(6.16)}\\
   \Tilde{\mathcal{O}}_{4\mu\nu}(t,x)&\equiv
   \delta_{\mu\nu}
   \mathring{\Bar{\chi}}(t,x)
   \overleftrightarrow{\Slash{D}}
   \mathring{\chi}(t,x),
\label{eq:(6.17)}\\
   \Tilde{\mathcal{O}}_{5\mu\nu}(t,x)&\equiv
   \delta_{\mu\nu}
   m\mathring{\Bar{\chi}}(t,x)
   \mathring{\chi}(t,x).
\label{eq:(6.18)}
\end{align}
The coefficients are given by, through the one-loop perturbation
theory,\footnote{The quadratic Casimirs are defined from anti-hermitian group
generators by~$\tr_R(T^aT^b)=-T(R)\delta^{ab}$, $T^aT^a=-C_2(R)1$,
and~$f^{acd}f^{bcd}=C_2(G)\delta^{ab}$, where $[T^a,T^b]=f^{abc}T^c$.}
\begin{align}
   c_1(t)&
   =\frac{1}{\Bar{g}(1/\sqrt{8t})^2}
   -b_0\ln\pi-\frac{1}{(4\pi)^2}
   \left[\frac{7}{3}C_2(G)
   -\frac{3}{2}T(R)N_f\right],
\label{eq:(6.19)}\\
   c_2(t)&
   =\frac{1}{8}
   \frac{1}{(4\pi)^2}
   \left[\frac{11}{3}C_2(G)
   +\frac{11}{3}T(R)N_f\right],
\label{eq:(6.20)}\\
   c_3(t)&
   =\frac{1}{4}\left\{1+\frac{\Bar{g}(1/\sqrt{8t})^2}{(4\pi)^2}C_2(R)
   \left[\frac{3}{2}+\ln(432)\right]\right\},
\label{eq:(6.21)}\\
   c_4(t)&=\frac{1}{8}d_0\Bar{g}(1/\sqrt{8t})^2,
\label{eq:(6.22)}\\
   c_5(t)&=-\frac{\Bar{m}(1/\sqrt{8t})}{m}
   \left\{1+\frac{\Bar{g}(1/\sqrt{8t})^2}{(4\pi)^2}C_2(R)
   \left[3\ln\pi+\frac{7}{2}+\ln(432)\right]\right\},
\label{eq:(6.23)}
\end{align}
and
\begin{equation}
   b_0=\frac{1}{(4\pi)^2}
   \left[\frac{11}{3}C_2(G)-\frac{4}{3}T(R)N_f\right],\qquad
   d_0=\frac{1}{(4\pi)^2}6C_2(R),
\label{eq:(6.24)}
\end{equation}
where $\Bar{g}(q)$ and $\Bar{m}(q)$ are the running gauge coupling and the
running mass in the $\text{MS}$ scheme, respectively. The validity of the
formula~\eqref{eq:(6.13)} for the lattice regularization has been examined
numerically for the thermodynamics of the $SU(3)$ pure Yang--Mills
theory~\cite{Asakawa:2013laa,Kitazawa:2014uxa,Itou:2015gxx,Kitazawa:2015ewp,%
Kitazawa:2016dsl} and of the $N_f=2+1$ QCD~\cite{Taniguchi:2016ofw,%
Kanaya:2016rkt} with very encouraging results. For the $SU(3)$ pure Yang--Mills
theory, two-point correlation functions of EMT has been
computed~\cite{Kitazawa:2016,FlowQCD} which even indicate the conservation law
of EMT.

Our universal formula~\eqref{eq:(6.13)} with the universal
coefficients~\eqref{eq:(6.19)}--\eqref{eq:(6.23)} is thought to be valid for
any sound regularization, including the lattice regularization with any
sensible lattice discretization. Moreover, from the dimensional counting, it is
conceivable that the expressions are valid even in
\emph{curved space\/}\footnote{If the scalar field exists, we will have
additional terms in EMT which contain the scalar curvature.} if straightforward
suitable modifications by using the curved (euclidean) metric are made. For
example, the flow equation~\eqref{eq:(5.1)} will be replaced by
\begin{equation}
   \partial_tB_\mu(t,x)=g^{\nu\rho}(x)\mathcal{D}_\nu G_{\rho\mu}(t,x),\qquad
   B_\mu(t=0,x)=A_\mu(x),
\label{eq:(6.25)}
\end{equation}
where $g^{\nu\rho}(x)$ and $\mathcal{D}_\nu$ are the curved space metric and
the covariant derivative, respectively. Thus, if we have a lattice
regularization on a curved manifold with which the general coordinate
invariance can be restored in the continuum limit (see
Refs.~\cite{Brower:2016vsl,Brower} for a recent attempt), we will be able to
use our universal formula to study physics related to EMT on a curved
manifold.\footnote{I would like to thank George T. Fleming for a discussion on
this possibility.}

Now, although our formula is expected to be universal, this universality holds
only after the renormalization and sending the cutoff to infinity. With the
lattice regularization, the universality hence holds only in the continuum
limit; we have to first set $a\to0$ while $t$ is kept fixed in physical unit
and then take the $t\to0$ limit
as~Eq.~\eqref{eq:(6.13)}.\footnote{In~Ref.~\cite{Kitazawa:2016dsl}, this double
limit is literally taken and very encouraging results are obtained.} In actual
numerical simulations with a finite lattice spacing~$a$, we thus have a natural
window for a sensible range of~$t$:
\begin{equation}
   a\ll\sqrt{8t}\ll\frac{1}{\Lambda},
\label{eq:(6.26)}
\end{equation}
where $\Lambda$ is a mass scale of the low energy physics (such as the hadron
mass, temperature, volume etc.). This picture comes from the fact that the
diffusion length of the flow equations is $\sim\sqrt{8t}$. Thus, when only
finite lattice spacings are available, we cannot simply set $t\to0$ and then it
is not a priori obvious whether our universal formula~\eqref{eq:(6.13)} is
practically useful. The numerical experiments so far indicate that the formula
is of real use, but still we have to understand and reduce the systematic error
associated with the $t\to0$ extrapolation. In the next two subsections, I
present some observations related to this issue.

\subsection{$O(t)$ correction in the continuum limit}

To make the extrapolation to~$t\to0$ for~Eq.~\eqref{eq:(6.13)} from very small
but finite~$t$, it will be quite helpful (or crucial) to have some idea on the
nature of the $O(t)$ term we have neglected from the small flow time expansion
to derive Eq.~\eqref{eq:(6.13)}, such as the last term
of~Eq.~\eqref{eq:(6.12)}.

From the general form of the small flow time expansion
in~Ref.~\cite{Luscher:2011bx}, such an $O(t)$ term has the structure,
\begin{equation}
   t\sum_ic_i(t;g,m;\mu)\left\{\mathcal{O}_i\right\}_R(x)
   =t\sum_ic_i(t;g,m;\mu)\sum_j\left(Z^{-1}\right)_{ij}\mathcal{O}_j(x),
\label{eq:(6.27)}
\end{equation}
where $c_i(t;g,m;\mu)$ are dimensionless coefficients and
$\{\mathcal{O}_i\}_R(x)$ are dimension~$6$ renormalized operators ($\mu$
denotes the renormalization scale). In the right-hand side, we have
re-expressed the renormalized operators in terms of bare
operators~$\mathcal{O}_j(x)$ by using the renormalization constants. Since we
are considering a composite operator of flowed fields which are some
combination of \emph{bare\/} fields, the derivative of~Eq.~\eqref{eq:(6.27)}
with respect to the renormalization scale~$\mu$ vanishes. Thus
\begin{equation}
   \left(\mu\frac{\partial}{\partial\mu}\right)_0
   \left[\sum_ic_i(t;g,m;\mu)\left(Z^{-1}\right)_{ij}\right]=0,
\label{eq:(6.28)}
\end{equation}
where the subscript~$0$ implies that the derivative is taken with all bare
quantities are kept fixed and, in terms of conventional renormalization group
functions $\beta\equiv\left(\mu\frac{\partial}{\partial\mu}\right)_0g$
and~$\gamma_m\equiv-\left(\mu\frac{\partial}{\partial\mu}\right)_0\ln m$, we
have
\begin{equation}
   \left(\mu\frac{\partial}{\partial\mu}
   +\beta\frac{\partial}{\partial g}
   -\gamma_mm\frac{\partial}{\partial m}
   -\gamma_i\right)c_i(t;g,m;\mu)=0,\qquad\text{no sum over~$i$}.
\label{eq:(6.29)}
\end{equation}
Here, we have introduced the matrix $\gamma_{ij}\equiv
-\sum_k\left(\mu\frac{\partial}{\partial\mu}\right)_0
\left(Z^{-1}\right)_{jk}Z_{ki}$ and assumed an appropriate operator basis with
which $\gamma_{ij}$ is diagonal, $\gamma_{ij}=\gamma_i\delta_{ij}$. Then, by
analyzing the $t\to0$ behavior of the solution to this equation, we conclude
\begin{equation}
   c_i(t;g,m;\mu)=C_i
   \Bar{g}(1/\sqrt{8t})^{2[1+\gamma_{i0}/(2b_0)]}
   \left[1+O\left(\Bar{g}(1/\sqrt{8t})^2\right)\right]+O(t),
\label{eq:(6.30)}
\end{equation}
where $\gamma_{i0}$ denotes the one-loop coefficient of the anomalous
dimension~$\gamma_i(g)$, $\gamma_i=\gamma_{i0}g^2+O(g^4)$ (the one-loop
coefficient in~$\beta$, $b_0$, is given by~Eq.~\eqref{eq:(6.24)}). Thus,
for~$t\to0$, the $O(t)$ term~\eqref{eq:(6.27)} is dominated by a dimension~$6$
operator with the \emph{smallest\/} one-loop anomalous dimension~$\gamma_{i0}$.
Although it seems not easy to enumerate anomalous dimensions of all
dimension~$6$ operators which appear in the small flow time expansion relevant
for EMT, the above observation still might be useful in making a realizable
$t\to0$ extrapolation in future numerical simulations.

\subsection{$O(a)$ correction for a fixed~$t$}

Quite often, I was asked the relationship between the conventional strategy for
EMT on the lattice in~Eq.~\eqref{eq:(3.10)} and the universal
formula~\eqref{eq:(6.13)} and, closely related to this point, why the one-loop
perturbative coefficients in~Eqs.~\eqref{eq:(6.19)}--\eqref{eq:(6.23)} are
expected or seem to work in numerical simulations, although the one-loop
perturbative renormalization constants have been known to be generally
insufficient for presently-accessible lattice parameters (for the axial vector
current, for example).\footnote{From colleagues including,
Shinya Aoki,
Shoji Hashimoto,
Naruhito Ishizuka,
Yoshio Kikukawa,
Yoshinobu Kuramashi,
and
Yusuke Taniguchi, to whom I would like to thank for discussions.}

I think that the point is quite related to an important property of the
gradient flow and deserves a closer look. Let us recall the representation of a
lattice composite operator in Symanzik's effective
theory~\cite{Symanzik:1983dc} (Ref.~\cite{Weisz:2010nr} is a very nice
exposition):
\begin{equation}
   Z_{\mathcal{O}}(a)\Hat{\mathcal{O}}(x)=\mathcal{O}(x)+a\mathcal{O}'(x)
   +a^2\mathcal{O}''(x)+\dotsb,
\label{eq:(6.31)}
\end{equation}
where $\Hat{\mathcal{O}}(x)$ is a \emph{bare\/} composite operator in lattice
theory with a spacing~$a$ and $Z_{\mathcal{O}}(a)$ is a renormalization constant
for~$\Hat{\mathcal{O}}(x)$; in the right-hand side, $\mathcal{O}(x)$,
$\mathcal{O}'(x)$, \dots, are all \emph{renormalized\/} composite operators in
Symanzik's effective theory. We see that only when the renormalization
constant~$Z_{\mathcal{O}}(a)$ is taken into account, the approach of the lattice
operator~$Z_{\mathcal{O}}(a)\Hat{\mathcal{O}}(x)$ to the continuum
one~$\mathcal{O}(x)$ becomes linear in~$a$.\footnote{In the pure Yang--Mills
theory, the approach would be linear in~$a^2$.} The lower order lattice
perturbation theory for~$Z_{\mathcal{O}}(a)$ would not be realizable because of
the ``tadpole dominance''~\cite{Lepage:1992xa} and a non-perturbative
determination of~$Z_{\mathcal{O}}(a)$ would be required. Now, if we apply the
same idea to a composite operator of flowed lattice fields, we would have
\begin{equation}
   \Hat{\mathcal{O}}(t,x)=\mathcal{O}(t,x)+a\mathcal{O}'(t,x)
   +a^2\mathcal{O}''(t,x)+\dotsb,
\label{eq:(6.32)}
\end{equation}
which is a relation holding in the effective theory considered
in~Ref.~\cite{Ramos:2015baa} for the gradient flow in lattice theory. Here, the
crucial difference from~Eq.~\eqref{eq:(6.31)} is that the left-hand side is
already a renormalized quantity and we do not need a renormalization constant
such as~$Z_{\mathcal{O}}(a)$ in~Eq.~\eqref{eq:(6.31)}.

Thus, from the representation~\eqref{eq:(6.32)}, we see that, \emph{for a fixed
flow time~$t>0$, the approach of the lattice operator $\Hat{\mathcal{O}}(t,x)$
to the continuum counterpart $\mathcal{O}(t,x)$ is linear in~$a$ (up to
logarithmic factors)}.\footnote{Again, in the pure Yang--Mills theory, the
approach would be linear in~$a^2$.} This expected $a\to0$ behavior of composite
operators of flowed fields can be found, for example, in Figs.~3 and~4
of~Ref.~\cite{Kitazawa:2016dsl}. Then, to reduce the slope to the $a\to0$
extrapolation, the improvement ideas~\cite{Ramos:2015baa,Fodor:2014cpa,%
Kamata:2016any} will be very useful.

The one-loop perturbative matching coefficients
in~Eqs.~\eqref{eq:(6.19)}--\eqref{eq:(6.23)}, which tell how we should take the
$t\to0$ extrapolation, are on the other hand obtained by the continuum
perturbation theory; this does not suffer from the tadpole dominance.

\section{Conclusion}

There have been rapid developments recently on the construction of EMT in
lattice field theory, an old but important problem, with encouraging results.
So far, tests and/or applications of new ideas are limited mostly to bulk
thermodynamics (i.e., one-point functions of EMT). Considering the vast amount
of potential applications, such as the spin/momentum structure of hadrons,
(quasi-)conformal field theory, large anomalous dimensions, gravity, etc.\ and
that they are mainly related to correlation functions of EMT, we expect much to
be explored.

On the small flow-time approach presented in the last section, it is
interesting that one can have a closed universal expression for EMT.
Applications to the bulk thermodynamics show encouraging results. In this
approach, as a ``by-product'' of the smearing effect of the flow, there is
a tendency that the noise in correlation functions is suppressed. Still, we
need to further understand and reduce the systematic error associated with the
$t\to0$ extrapolation.

\section*{Acknowledgments}
I would like to thank
Francesco Capponi,
Leonardo Giusti,
Agostino Patella,
Michele Pepe,
and
Antonio Rago
for providing me useful information on their works.
I am also grateful to
collaborators in
the FlowQCD Collaboration
and
the WHOT-QCD Collaboration
for enjoyable collaborations.
The work of H.~S. is supported in part by JSPS Grants-in-Aid for Scientific
Research Grant Number~16H03982.


\end{document}